\documentstyle[hebrew,psfig,prb,aps,multicol]{revtex}

\tighten

\begin{document}
\def\H{\mbox{$\cal H$}}
\def\scr#1{\mbox{\scriptsize #1}}
\def\mb#1{\mbox{\boldmath $#1$}}
\def\ds{\displaystyle}
%\draft
\title{Braggoriton--Excitation in Photonic Crystal Infiltrated with Polarizable Medium}
\author{A. Yu. Sivachenko, M. E. Raikh, and Z. V. Vardeny}
\address{Department of Physics, University of Utah, Salt Lake City, Utah
84112}
\date{\today}
\maketitle
\begin{abstract}
Light propagation in a photonic crystal infiltrated
with polarizable molecules is considered. We demonstrate 
that the interplay between the spatial dispersion caused by 
Bragg diffraction and polaritonic frequency dispersion
gives rise to novel propagating
excitations, or {\em braggoritons}, with intragap frequencies. 
We derive the braggoriton dispersion
relation and show that it is governed by two parameters,
namely, the strength of light-matter interaction and 
detuning between the
Bragg frequency and that of the infiltrated molecules.
We also study defect-induced states when 
the photonic band gap is divided into
two subgaps by the braggoritonic branches and find that 
each defect creates {\em two} intragap localized states
inside {\em each} subgap.
\end{abstract}

\begin{multicols}{2}

\section{Introduction}

Photonic crystals and in particular
photonic band gap (PBG) materials\cite{PBG,PBGJohn} have recently
attracted much attention\cite{soukoulis,Joannopoulos} due to their
rich physics and possible applications. In these systems the
dielectric function is periodically modulated and, as a result,
their optical properties are dominated by light diffraction effects.  When
Bragg diffraction conditions are met then light scattering is very
strong, so that within certain frequency intervals near the
resonances light propagation is inhibited.
%Depending on structure of a sample,
%the spectrum of photons exhibits  {\em forbidden gaps} in some directio%ns 
%(incomplete band gap, or pseudogap), e.g. for 2D array of dielectric
%rods\cite{platzband}) or for 3D fcc lattice of dielectric 
%spheres\cite{fcc-band,soukoulis-band}, or in all directions
%(complete band gap), as for 3D diamond lattice\cite{soukoulis-band}.

Since the subject of photonic crystals 
was intro\-duced,\cite{PBG,PBGJohn} one of the main 
goals of
photonic band-structure calculations was to engineer structures
with a {\em complete}
band gap, {\em i.e.} with no propagating solutions of Maxwell's equations
within a certain {\em forbidden gap}. The pursuit of this goal has
generated a stream of studies that are too numerous to be cited
here; early works are reviewed in
Refs.\onlinecite{soukoulis,Joannopoulos}. Here we only mention that
a complete band gap in two dimensions (2D) was theoretically
predicted\cite{platzband,maradudin} and experimentally 
demonstrated\cite{platzband} for an array of dielectric rods. In the
quest for a structure having a complete PBG in three dimensions 
(3D), the diamond
lattice was shown\cite{soukoulis-band} to be more promising than a
simple face centered cubic (fcc) lattice\cite{soukoulis-band,fcc-band}.

The frequency gap in the photonic spectrum sets a stage for a 
number of physical
effects. The prime effect, namely the inhibition of 
spontaneous emission for an emitter
with transition frequency within the gap, was already suggested
in the pioneering works.\cite{PBG,JohnLoc1,spont} Furthermore, since
light cannot leave the emitting atom, a coupled atom-field
in-gap state is formed, in which the atomic level is ``dressed'' by its own
exponentially localized radiation field.\cite{JohnLoc1,JohnLoc}  It
was also demonstrated that although a single photon cannot propagate
inside the gap, nevertheless a non-linear medium 
embedded inside the photonic crystal
gives rise to multi-photon bound states\cite{multiph}, or gap
solitons\cite{solitons} that result in self-induced transparency. Yet
another consequence of PBG is the modification of cooperative emission
with frequency close to the band edge. In particular PBG was
shown to change the rate of superradiant emission from an ensemble
of emitters.\cite{spont,Johnsuper} Lastly, PBG structures facilitate
strong Anderson localization of photons\cite{genack} because the
sharp density of states within the gap spectral range
necessitates a reinterpretation of the Ioffe-Regel
criterion.\cite{PBGJohn}

PBG structures with a defect constitute 
a separate area of study initiated by the classical
works in Refs.\onlinecite{platzband,joanndefect}. 
These structures are important since the defects cause
localized intragap states. For these states, the PBG sample acts
as a resonator with a very high quality factor. This property was recently
used for designing a low-threshold PBG defect-mode laser.\cite{bglaser}

Another class of materials with a forbidden gap for light propagation
are spatially homogeneous, but frequency-dispersive media. The energy gap
in these systems has a polaritonic origin, {\em i.e.} it is formed
due to the interaction of light with the medium polarization.\cite{kittel}
This energy gap can be viewed as the result of anticrossing between the 
photonic and excitonic dispersion relation branches.  
Some non-trivial manifestations of the
polaritonic gap were recently explored in
Refs.\onlinecite{RupasovGap,RupasovGap1}. In
these papers a general model of two-level systems interacting with
elementary electromagnetic excitations with a gap in the spectrum was
solved by means of the Bethe ansatz technique. Within this model a
very rich excitation spectrum was found,\cite{RupasovGap,RupasovGap1} 
consisting of ordinary
solitons, single-particle impurity bound states and massive pairs of
confined gap excitations and their bound complexes --- dissipationless
quantum gap solitons.

%The model of Refs.\cite{RupasovGap,RupasovGap1,RupasovGap2,RupasovSIT} 
%is quite general and can be applied to any system with a gap in the pho%tonic
%spectrum. 

%The gap is, however, inavoidably
%{\em either} of the Bragg {\em or} of polaritonic origin.

% Most photonic crystals do not have complete band gap...

Most of the available photonic crystals nowadays however
have {\em incomplete}
PBGs; this means that light propagation is forbidden only along
certain directions inside the crystal. 
A prominent example are opals, representing
self-assembled monodispersed silica balls\cite{opals}
arranged in a fcc type lattice. 
Although opals have only
incomplete PBG, the voids between the balls
 can be infiltrated by various
media, which brings about non-trivial physics.  In particular, the
medium may contain polarizable molecules. Infiltrated opal with
polarizable molecules combines therefore
polaritonic and Bragg-diffractive properties. Obviously, both effects
coexist independently when the Bragg ($\omega=\omega_{\scr{B}}$) and
polaritonic ($\omega=\omega_{\scr{T}}$) resonances are well separated in
frequency. A completely different situation occurs when
$\omega_{\scr{B}}\approx\omega_{\scr{T}}$.  This may be easily achieved in 
infiltrated opals that gives rise to
a peculiar {\em interplay} between various frequency
dispersions. This interplay is the subject of the present paper.

Our most important finding pertains to the case when the polaritonic gap
of the polarized molecules infiltrating the opal lies
within the opal PBG. We demonstrate that such an overlap gives rise
to novel massive {\em propagating} excitations having frequencies
inside the Bragg gap, that we dubbed here 
{\em braggoritons}. 
In other words, the Bragg gap {\em splits} into
two sub-gaps, so that the braggoritonic branches are isolated from the rest
of the spectrum. We found that the braggoriton dispersion relation 
is very sensitive to
the frequency detuning between $\omega_{\scr{B}}$ 
and $\omega_{\scr{T}}$ and to the relative
width of the polaritonic gap (or, alternatively
Rabi frequency) and the Bragg gap.

The principal assumption we adopt here
is that the Bragg gap, $\Delta\omega_{\scr{B}}$, is narrow compared 
to $\omega_{\scr{B}}$; this is actually the
case in opals. The small value of $\Delta\omega_{\scr{B}}/\omega_{\scr{B}}$
($\ll1$)
enables us then to obtain analytical results.
In addition we also study the phase slip related intragap defect states for
$\omega_{\scr{B}}\approx\omega_{\scr{T}}$. 
In the absence of polaritonic effect, the
underlying physics of the defect-induced intragap states was already
discussed in the original PBG paper\cite{PBG}. An analogy was drawn between
a defect state and a localized mode in a distributed feedback resonator,
that originates from a phase slip. We extend this picture to incorporate
polarizable medium and show that when the Bragg gap splits into two
sub-gaps, then an existing phase slip gives rise to {\em two} 
localized states with frequencies within {\em each} of the sub-gaps.
 
Our paper is organized as follows: In order to introduce
the notations, we separately review in Sec.~II the derivation of the PBG 
and polaritonic spectra using the second quantization representation. In
Sec.~III we consider the combined Hamiltonian in the second
quantization representation and
diagonalize it by a unitary transformation. This yields the dispersion
relations
for the two excitations outside the gap, or Bloch-like waves,   
and the two intragap branches, or braggoritons.
 The properties of these novel excitations are
analyzed in Sec.~IV.  We use them in Sec.~V to determine the intragap
frequencies
of the defect-induced localized states.  Concluding remarks are presented
in Sec.~VI.

\section{Second quantized PBG and polaritonic Hamiltonians}

The Hamiltonian \H\ of the system under study is the sum of three terms

%We start with the Hamiltonian
%
\begin{equation} \label{H}
\H=\H_{\scr{ph}}+\H_{\scr{m}}+\H_{\scr{m-ph}}.
\end{equation}
%
%where the terms on the right hand side of~(\ref{H}) are the Hamiltonian%s for
%photons in the photonic crystal, for polarazible medium, and for photon%-medium 
%interaction, correspondingly. 
The first term, $\H_{\scr{ph}}$, describes the photons in a
photonic crystal. The second term, $\H_{\scr{m}}$,
is the Hamiltonian of the polarizable medium; $\H_{\scr{m-ph}}$ describes
the photon-medium coupling. In this Section we review two
limiting cases: ({\em i}) no polarizable medium  
($\H_{\scr{m}}\equiv 0$),
and ({\em ii}) no modulation of the dielectric constant.

\subsection{Incomplete PBG}
The general form of the Hamiltonian $\H_{\scr{ph}}$ is
\begin{equation} \label{Hem}
\H_{\scr{ph}}=\frac{1}{8\pi}\int d\mb{r}\;  \left[ \varepsilon(\mb{r})\mb{E}^2+
\mb{H}^2\right],
\end{equation}
where $\mb{E}$ and $\mb{H}$ are respectively 
the electric and magnetic fields.
For a constant dielectric function, $\varepsilon(\mb{r})\equiv
\varepsilon_0$, the second quantized form\cite{Landau} of
the Hamiltonian~(\ref{Hem}) reduces to a sum over 
oscillators representing plane waves with frequencies 
$\omega_k=ck/\sqrt{\varepsilon_0}$, where
$k$ is the wave vector. Modulation of 
$\varepsilon(\mb{r})$ causes light diffraction, so that the
plane wave solutions are no longer the correct eigenfunctions 
of the Hamiltonian~(\ref{Hem}). 
Below we consider a photonic crystal with an incomplete
PBG along the $z$ axis.
This situation can be approximated by a one-dimensional modulation of 
$\varepsilon(\mb{r})$ along the $z$ direction. If the Bragg gap is 
relatively narrow as assumed above, then it may be sufficient 
to consider only the first harmonics in $\varepsilon(z)$:
\begin{equation}\label{epsilon}
\varepsilon(z)=\varepsilon_0+\delta\varepsilon\cos(\sigma z+\phi).
\end{equation}
Here $\delta\varepsilon$ ($\ll\varepsilon_0$) 
is the modulation amplitude, $\sigma=2\pi/d$, where
$d$ is the modulation period and $\phi$ is the dielectric modulation
phase.
We assume for simplicity that the electromagnetic field 
propagates along the $z$-direction and is homogeneous 
in the $xy$ plane. In this case light polarization
is irrelevant. Generalization to 
 arbitrary propagation direction
is straightforward.  
The Fourier components of $\varepsilon(z)$ in~(\ref{epsilon})
couple the original photon oscillators with  momenta 
$k$ and  $k\pm\sigma$.
These coupled
oscillators form an infinite series that is
constructed by successive addition 
(subtraction) of $\sigma$. However, if 
$\delta\varepsilon\ll\varepsilon_0$ 
and the wavevectors domain is 
restricted to the vicinity of the first Bragg
resonance at $k\approx\sigma/2$, then the Hamiltonian~(\ref{Hem})
can be truncated.
In this case only the coupling to the near-resonance
backscattered photons with momenta $(\sigma-k)\approx\sigma/2$ must be 
retained, so that the Hamiltonian~(\ref{H}) takes the form
\begin{eqnarray}\nonumber
& &\H_{\scr{ph}}= \\ \nonumber
& &\sum_{q}\left\{\omega(q) \hat{a}^+_{\rightarrow}(q) 
\hat{a}^{}_{\rightarrow}(q)+\omega(-q)\hat{a}^+_{\leftarrow}(-q) 
\hat{a}^{}_{\leftarrow}(-q)\right. \\ \label{Hph}
& &+\left. \Omega_{\scr{B}}\left[
e^{i\phi}\;\hat{a}^+_{\rightarrow}(q)\hat{a}^{}_{\leftarrow}(-q)+
e^{-i\phi}\;\hat{a}^+_{\leftarrow}(-q)\hat{a}^{}_{\rightarrow}(q)\right]\right\}.
\end{eqnarray}
Here, we introduced the notations: $q=k-\sigma/2$, $\hat{a}_{\rightarrow}(q)=
\hat{a}_k$ and $\hat{a}_{\leftarrow}(-q)=
\hat{a}_{k-\sigma}$ for $k\approx\sigma/2$, where $\hat{a}_k$ is 
the usual photon annihilation
operator. In the notations introduced in~(\ref{Hph}), 
the frequencies of the photonic 
branches are given by:
\begin{equation}
\omega(q)=\frac{c(q+\sigma/2)}{\sqrt{\varepsilon_0}}=
\omega_{\scr{B}}\left(1+\frac{2q}{\sigma}\right),
\end{equation}
where $\omega_{\scr{B}}=c\sigma/(2\sqrt{\varepsilon_0})$ is the 
Bragg frequency. We define the coupling constant, $\Omega_{\scr{B}}$, as
the half-width of the Bragg gap, {\em i.e.} 
$\Omega_{\scr{B}}=\frac{1}{2}\Delta\omega_{\scr{B}}$. It can be 
shown that $\Omega_{\scr{B}}$ is related to 
the amplitude of the 
dielectric function modulation:
$\Omega_{\scr{B}}=\omega_{\scr{B}}\delta\varepsilon/(2\varepsilon_0)$. 
The summation in Eq.~(\ref{Hph}) is performed over the $k$ domain
$|q|\ll\sigma/2$.

It is straightforward to diagonalize the Hamiltonian in Eq.~(\ref{Hph})  
with the use of the following unitary transformation 
\begin{equation} \begin{array}{c} 
\hat{a}_{\rightarrow}(q)=\cos\theta\;\hat{\beta}_1(q)+\sin\theta\;e^{i\phi}\;
\hat{\beta}_2(q), \\
\hat{a}_{\leftarrow}(-q)=-\sin\theta\;e^{-i\phi}\;\hat{\beta}_1(q)+\cos\theta\;
\hat{\beta}_2(q),\end{array}
\end{equation}
where
\begin{equation} \label{costheta}
\cos2\theta=\frac{\omega(q)-\omega(-q)}{\sqrt{\left[\omega(q)-
\omega(-q)\right]^2+4\Omega_{\scr{B}}^2}}.
\end{equation}
The new operators $\hat{B}_1$ and $\hat{B}_2$ describe the creation
(annihilation) of pairs of Bloch waves, 
that consist of
forward and backscattered photons near the Bragg frequency. The 
diagonalized Hamiltonian~(\ref{Hph}) takes the form
\begin{equation}
\H_{\scr{ph}}=\sum_{q}\omega^{(1)}_{\scr{B}}(q)\hat{\beta}^+_1(q)
\hat{\beta}_1(q)
+\omega^{(2)}_{\scr{B}}(q)\hat{\beta}^+_2(q)\hat{\beta}_2(q),
\end{equation}
where the dispersion relations of the two photonic branches are given by: 
\begin{eqnarray} \nonumber 
& &\omega^{(1,2)}_{\scr{B}}(q)=\\ \nonumber
& &\frac{1}{2}\Biggl[\omega(q)+\omega(-q)\pm
\sqrt{(\omega(q)-\omega(-q))^2+4\Omega^2_{\scr{B}}}\;\Biggr]= \\ \label{ph-br}
& & \omega_{\scr{B}}\pm\sqrt{\left(
\frac{2\omega_{\scr{B}}}{\sigma}\right)^2q^2+\Omega_{\scr{B}}^2}\,.
\end{eqnarray}
As mentioned above, the width of the PBG from Eq.~(\ref{ph-br})
is $2\Omega_{\scr{B}}$.

\subsection{Polarizable medium}

The Hamiltonians $\H_{\scr{m}}$ of  polarizable medium  and 
$\H_{\scr{m-ph}}$ of light-polarization interaction in Eq.~(\ref{H}) can be 
written in the second quantization form as:
\begin{eqnarray} \nonumber
\H_{\scr{m}}+\H_{\scr{m-ph}}& = &
\omega_{\scr{T}}\sum_k \hat{b}_k^+\hat{b}_k^{} \\
& + & \Omega_{\scr{P}}\sum_k\left[\hat{b}_k^+\hat{a}_k^{}+
\hat{a}_k^{+}\hat{b}_k^{}\right],
\end{eqnarray}
where $\hat{b}$ is the annihilation operator of the medium 
 excitations 
({\em e.g.} excitons or optical phonons), which
we assume here to be dispersionless
having frequency $\omega_{\scr{T}}$; $\Omega_{\scr{P}}$
denotes the light-medium coupling strength that is proportional
to the Rabi frequency.
In the absence of the Bragg scattering term ($\Omega_{\scr{B}}=0$)  
the complete Hamiltonian~(\ref{H})reduces to
the conventional polaritonic Hamiltonian  $\H_{\scr{pol}}$, given by:
\begin{eqnarray}\nonumber
\H_{\scr{pol}}&=&\sum_k\omega_k\hat{a}^+_k\hat{a}^{}_k+
\omega_{\scr{T}}\sum_k \hat{b}_k^+\hat{b}_k^{} \\ \label{polar}
&+&\Omega_{\scr{P}}
\sum_k\left[\hat{b}_k^+\hat{a}_k^{}+\hat{a}_k^{+}\hat{b}_k\right],
\end{eqnarray}
with eigenstates representing the mixture of light and medium excitations 
\begin{equation} \begin{array}{c}
\hat{a}_k=\cos\psi\;\hat{\pi}_1(k)+\sin\psi\;\hat{\pi}_2(k), \\
\hat{b}_k=-\sin\psi\;\hat{\pi}_1(k)+\cos\psi\;\hat{\pi}_2(k),
\end{array}
\end{equation}
where
\begin{equation} \label{cospsi}
\cos2\psi=\frac{\omega_k-\omega_{\scr{T}}}{\sqrt{(\omega_k-\omega_{\scr{T}})^2+
4\Omega_{\scr{P}}^2}}, \hskip2mm\mbox{($\omega_k=ck/\sqrt{\varepsilon_0}\;$)}.
\end{equation}
With the new operators $\hat{\pi}_1$ and $\hat{\pi}_2$, the 
Hamiltonian  (\ref{polar}) is diagonalized:
\begin{eqnarray} \nonumber
\H_{\scr{pol}}& =&\sum_k\omega^{(1)}_{\scr{P}}(k)
\hat{\pi}_1^+(k)\hat{\pi}_1^{}(k) \\
&+&\sum_k\omega^{(2)}_{\scr{P}}(k)\hat{\pi}_2^+(k)\hat{\pi}_2^{}(k),
\end{eqnarray}
where the frequencies of the polaritonic branches are
given by
\begin{equation} \label{pol-br}
\omega^{(1,2)}_{\scr{P}}(k)=\frac{1}{2}\Biggl[\omega_k+\omega_{\scr{T}}\pm
\sqrt{(\omega_k-\omega_{\scr{T}})^2+4\Omega_{\scr{P}}^2}\;\Biggr].
\end{equation}
%
%From comparison with the spectrum of polaritons it is clear that the 
%coupling 
%constant is $\Omega_P=\sqrt{\omega_T\omega_{LT}}$, where 
The Rabi splitting at resonance, {\em i.e.} at 
$\omega_k=\omega_{\scr{T}}$ is $2\Omega_{\scr{P}}$.
Eq.~(\ref{pol-br}) allows to express the phenomenological
parameter $\Omega_{\scr{P}}$ through the observables. Namely,
$\Omega_{\scr{P}}=\sqrt{\omega_{\scr{T}}\omega_{\scr{LT}}/2\;}$, where
$\omega_{\scr{LT}}\ll\omega_{\scr{T}}$
is the transverse-longitudinal splitting.

We note that the above description is valid only for wavenumbers
$k$ in the vicinity
of ``crossing'' of the excitation branches, 
where  $\omega_k\simeq\omega_{\scr{T}}$. It does not capture, however, the
correct behavior\cite{kittel} of the polaritonic branches for 
$k\rightarrow 0$. In this limit an additional term of the type
$a_kb_{-k}+\mbox{c.c.}$ should be taken into account in Eq.~(\ref{polar}).

\section{Diagonalization of the full Ha\-mil\-tonian}

Now let us consider the full 
Hamiltonian~(\ref{H}) with both Bragg scattering
and light-medium interaction included,
\end{multicols}
\parbox{\textwidth}{
\begin{multicols}{2}
\begin{eqnarray}\label{full} \nonumber
\H&=&\sum_{q}\left[\omega(q)\hat{a}^+_{\rightarrow}(q)
\hat{a}^{}_{\rightarrow}(q)+
\omega(-q)\hat{a}^+_{\leftarrow}(-q)\hat{a}^{}_{\leftarrow}(-q)
\right]\\ \nonumber
&+&\Omega_{\scr{B}}\sum_{q}\left[ e^{i\phi}\;
\hat{a}^+_{\rightarrow}(q)\hat{a}^{}_{\leftarrow}(-q)+
e^{-i\phi}\;\hat{a}^{+}_{\leftarrow}(-q)\hat{a}^{}_{\rightarrow}(q)\right] \\
\nonumber
&+&\omega_{\scr{T}}
\sum_{q} \left[\hat{b}_{\rightarrow}^+(q)\hat{b}_{\rightarrow}^{}(q)
+\hat{b}_{\leftarrow}^+(-q)\hat{b}_{\leftarrow}(-q)^{}\right]\\ \nonumber
&+&\Omega_{\scr{P}}
\sum_{q}\left[ \hat{b}_{\rightarrow}^+(q)\hat{a}_{\rightarrow}^{}(q)+
\hat{a}_{\rightarrow}^{+}(q)\hat{b}^{}_{\rightarrow}(q)\right.\\
&+& \left.\hat{b}^+_{\leftarrow}(-q)
\hat{a}^{}_{\leftarrow}(-q)+\hat{a}^+_{\leftarrow}(-q)\hat{b}^{}_{\leftarrow}(-q)
\right],
\end{eqnarray}
where we have again truncated the ``Bragg'' Hamiltonian
in~(\ref{H}) by including only near-resonance
terms. If a column of operators $\hat{c}=\{\hat{a}_{\rightarrow}(q),\;
\hat{a}_{\leftarrow}(-q),\;\hat{b}_{\rightarrow}(q),\;\hat{b}_{\leftarrow}(-q)\}$
is introduced, then the Hamiltonian (\ref{full}) can be formally 
rewritten in a matrix 
form $\H=\hat{c}^+H\hat{c}$, where
\begin{equation}
 H=\left(
\begin{array}{cccc}
\omega(q) & \Omega_{\scr{B}}\;e^{i\phi} &\Omega_{\scr{P}} & 0\\
\Omega_{\scr{B}}\;e^{-i\phi} & \omega(-q) & 0 & \Omega_{\scr{P}} \\
\Omega_{\scr{P}} & 0 & \omega_{\scr{T}} & 0 \\
0 & \Omega_{\scr{P}} & 0 & \omega_{\scr{T}}
\end{array}\right).
\end{equation}
%
%We have to find a unitary transformation $\hat{c}=S\hat{C}$ to new operators
%$\hat{C}=\{\mb{\hat{\cal B}}_1,\;\hat{\mbox{\ta a}}_1,\;
%\hat{\mbox{\ta a}}_2,\;
%\mb{\hat{\cal B}}_2\}$
%that annihilate
%mixed light-matter states. The Hamiltonian~(\ref{full}) 
%in the new representation becomes
%%
%\begin{equation} \label{rotation}
%\H=\hat{C}^+S^+HS\hat{C},
%\end{equation}
%
%and we require the matrix $S^+HS$ to be diagonal.
%Hence the problem is reduced to diagonalization of the matrix $H$. 
The four $H$ eigenvalues yield  
the dispersion relations of the {\em four} excitation branches, whereas 
the eigenvectors determine the unitary
transformation diagonalizing $\H$. 
%Since
%the matrix $H$ is hermitian, then its eigenvectors are orthogonal and $S$
%is a unitary matrix that guarantees proper commutation relations for the
%new excitation creation/annihilation operators.

The characteristic equation for the eigenvalues $\omega$ of the matrix $H$ 
reads
\end{multicols}
}

\vskip-4mm

\vbox{\hbox to9.2cm{\hrulefill\vrule height2mm}}  
\begin{equation} \label{eigen}
\Biggl[\;\Bigl(\omega(q)-\omega\Bigr)
\Bigl(\omega(-q)-\omega\Bigr)-\Omega_{\scr{B}}^2-
\Biggl(\frac{\omega(q)+\omega(-q)-
2\omega}{\omega_{\scr{T}}-\omega}\Biggr)\;\Omega_{\scr{P}}^2\;\Biggr]\;
(\omega_{\scr{T}}-\omega)^2+\Omega_{\scr{P}}^4=0.
\end{equation}
If the light-matter coupling is absent, {\em i.e.} $\Omega_{\scr{P}}=0$, 
then the roots
of~(\ref{eigen}) reduce to two pure medium excitations
with unperturbed frequency
$\omega_{\scr{T}}$
propagating
in the forward and  backward directions along $z$,  and two purely 
photonic excitations with disperion relation given by Eq.~(\ref{ph-br}) that
results from the Bragg scattering. 
If, on the other hand, the Bragg scattering is 
absent, {\em i.e.} $\Omega_{\scr{B}}=0$, 
then the roots of Eq.~(\ref{eigen}) reduce
 to two pairs
of polariton branches with dispersion relation given by Eq.~(\ref{pol-br}).
%One of the pairs corresponds to forward,
%and the other to the backward propagating polaritons, which are not cou%pled
%in the absense of backscattering, $\Omega_B=0$.

It appears that the unitary transformation in {\em four-dimensional} space 
diagonalizing the Hamiltonian $\H$ can be parameterized by {\em two} 
angles:
\begin{equation} \label{totalrot}
\left(\begin{array}{c}
\hat{a}_{\rightarrow}(q) \\
\hat{a}_{\leftarrow}(-q) \\
\hat{b}_{\rightarrow}(q) \\
\hat{b}_{\leftarrow}(-q)
\end{array}\right)= 
 \left(
\begin{array}{cccc}
\cos\theta\;\cos\tilde{\psi} & \sin\theta\;\sin\tilde{\psi} &
\cos\theta\;\sin\tilde{\psi} & \sin\theta\;\cos\tilde{\psi} \\
\sin\theta\;\cos\tilde{\psi} & -\cos\theta\;\sin\tilde{\psi} &
\sin\theta\;\sin\tilde{\psi} & -\cos\theta\;\cos\tilde{\psi} \\
\cos\theta\;\sin\tilde{\psi} & \sin\theta\;\cos\tilde{\psi} &
-\cos\theta\;\cos\tilde{\psi} & -\sin\theta\;\sin\tilde{\psi} \\
\sin\theta\;\sin\tilde{\psi} & -\cos\theta\;\cos\tilde{\psi} &
-\sin\theta\;\cos\tilde{\psi} & \cos\theta\;\sin\tilde{\psi} 
\end{array}\right)\cdot 
 \left(
\begin{array}{c}
\hat{\mb{\cal B}}_2 \\
\hat{\mbox{\ta a}}_2 \\
\hat{\mbox{\ta a}}_1 \\
\hat{\mb{\cal B}}_1 
\end{array}\right),
\end{equation}
where $\hat{\mb{\cal B}}_1$, $\hat{\mbox{\ta a}}_1$,
$\hat{\mbox{\ta a}}_2$, and
$\hat{\mb{\cal B}}_2$ are new operators that annihilate mixed
light-matter states. The angle $\theta$ in Eq.~(\ref{totalrot}) is precisely
the ``Bragg'' rotation angle introduced in Eq.~(\ref{costheta}) [for 
simplicity we set $\phi=0$ for the modulation phase in 
Eq.~(\ref{totalrot})]. 
The second angle, $\tilde{\psi}$ is defined by the following relation
\begin{equation} \label{costildepsi} 
\cos2\tilde{\psi}=
\frac{
\sqrt{\;\left[\omega(q)-\omega(-q)\right]^2+4\Omega_{\scr{B}}^2}-
2(\omega_{\scr{T}}-\omega_{\scr{B}})
}{\sqrt{
\left\{2(\omega_{\scr{T}}-\omega_{\scr{B}})-
\sqrt{\;\left[\omega(q)-\omega(-q)\right]^2+4\Omega_{\scr{B}}^2}\;\right\}^2+
16\Omega_{\scr{P}}^2}}.
\end{equation}

\vskip-3mm

\vbox{\hspace*{8.8cm}\hbox to 8.7cm{\vrule depth2mm\hrulefill}}

\vspace*{-2mm}

\begin{multicols}{2}
Naturally, for $\Omega_{\scr{B}}=0$ the angle $\tilde{\psi}$
reduces to the polaritonic rotation angle $\psi$ in Eq.~(\ref{cospsi}). 
In the presence
of the Bragg scattering, however, this rotation angle 
also depends on the ``Bragg'' parameters $\omega_{\scr{B}}$ and 
$\Omega_{\scr{B}}$. Therefore it is the angle $\tilde{\psi}$ that 
characterizes the interplay between the polaritonic and diffraction effects.

\section{Braggoritonic excitations}

In order to analyze the solutions of  
Eq.~(\ref{eigen}), it is 
convenient to introduce the following dimensionless variables. 
We measure frequencies, $\Delta$, 
from the Bragg frequency, $\omega_{\scr{B}}$, and express them in 
units of the Bragg gap $2\Omega_{\scr{B}}$:
\begin{equation} \label{deltadef}
\Delta=\frac{\omega-\omega_{\scr{B}}}{2\Omega_{\scr{B}}}.
\end{equation}
In analogy with Eq.~(\ref{deltadef}) we introduce the dimensionless 
frequency detuning, $\delta$, 
of $\omega_{\scr{T}}$ from the Bragg 
frequency $\omega_{\scr{B}}$, where
\begin{equation}
\delta=\frac{\omega_{\scr{T}}-\omega_{\scr{B}}}{2\Omega_{\scr{B}}}.
\end{equation}
As seen from Eq.~(\ref{ph-br}), the natural unit for 
the wavevector deviation, $q$, 
from the Bragg wavevector, $\sigma/2$, is 
$\sigma\Omega_{\scr{B}}/\omega_{\scr{B}}$. 
Hence, we introduce the dimensionless parameter
\begin{equation}
Q=\left(\frac{\omega_{\scr{B}}}{\sigma\Omega_{\scr{B}}}\right)\;q.
\end{equation}
%
%%%%%%%%%%%%%%%%%%%%%%%%%%%%%%%%%%%%%%%%%%%%%%%%%%%%%%%%
%%%%%%%%%%%%%%%%%%%%%%%%%%%%%%%%%%%%%%%%%%%%%%%%%%%%%%%%
\parbox[t]{8.65cm}{
\psfig{figure=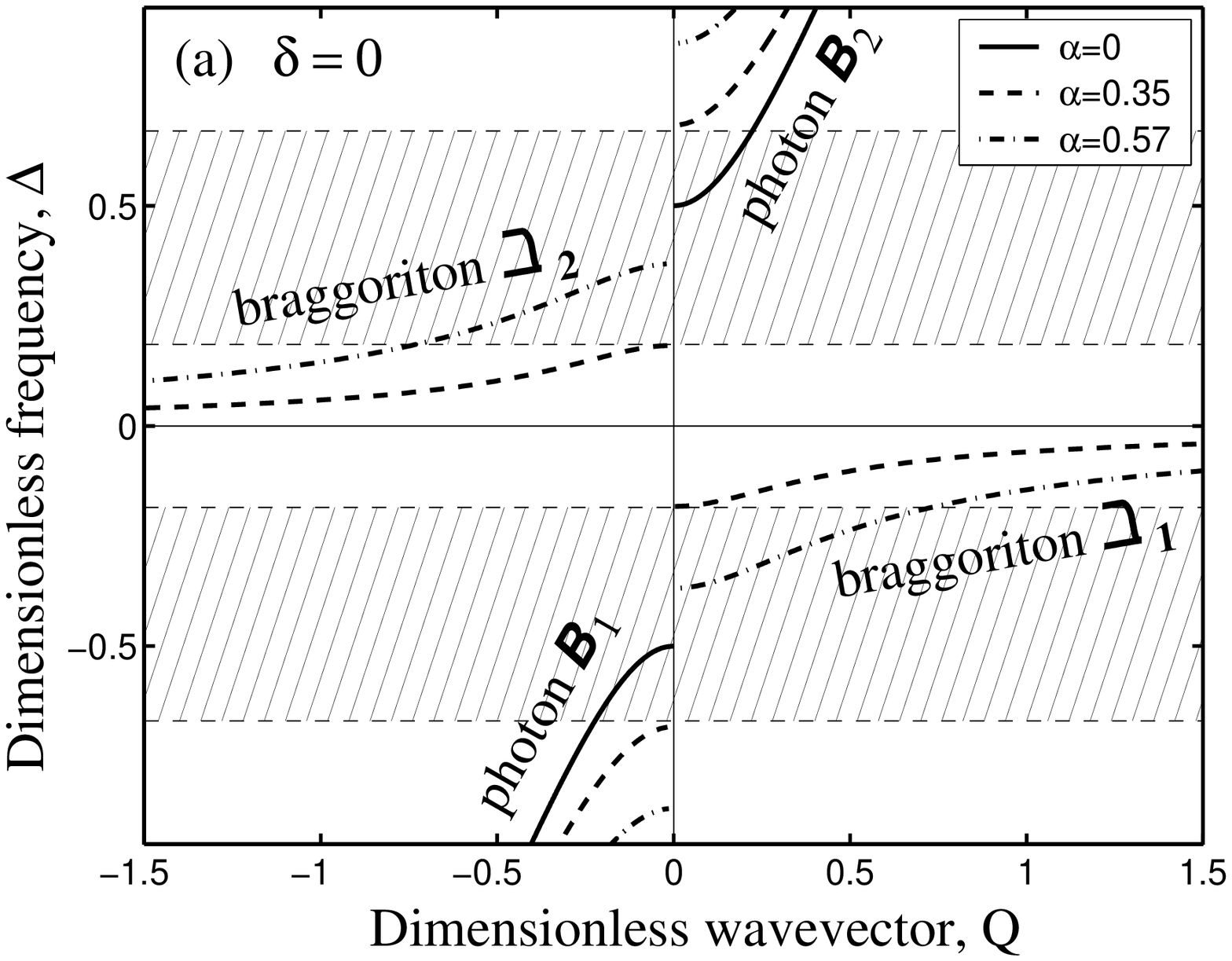,width=8.65cm}

\vskip2mm

\psfig{figure=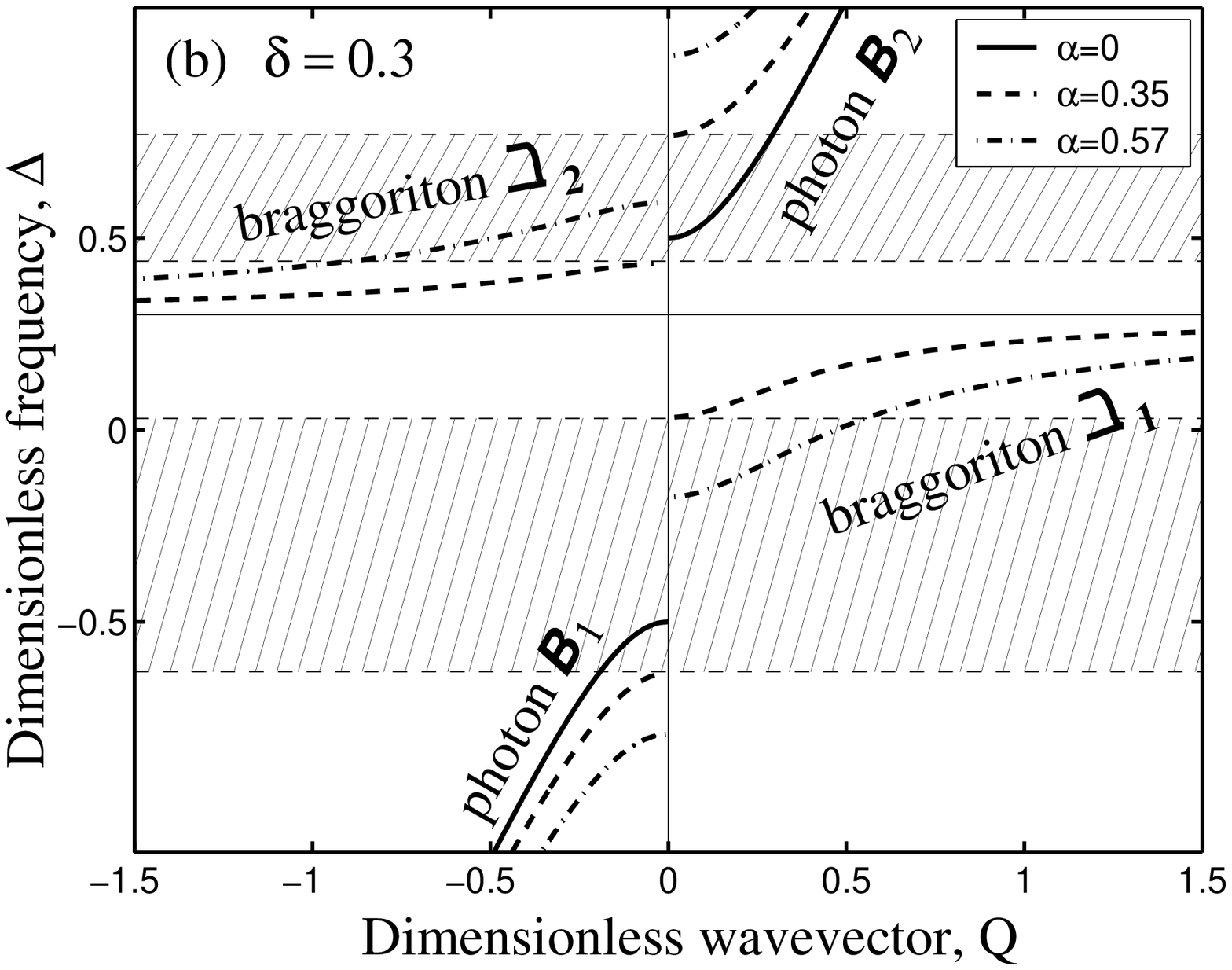,width=8.65cm}

\vskip2mm

{
\small{\bf Fig. 1} Dispersion of mixed photonic-medium excitations for
various coupling strengths: $\alpha=0, 0.35, 0.57$. (a) $\delta=0$; 
(b) $\delta=0.3$.
The shaded area represents the two forbidden sub-gaps at $\alpha=0.35$.
}

\vskip4mm

}
%%%%%%%%%%%%%%%%%%%%%%%%%%%%%%%%%%%%%%%%%%%%%%%%%%%%%%%%
%%%%%%%%%%%%%%%%%%%%%%%%%%%%%%%%%%%%%%%%%%%%%%%%%%%%%%%%
With the new notations, the excitation spectrum determined 
by Eq.~(\ref{eigen}) can be 
rewritten in a more concise form
\begin{equation} \label{spectrum}
\left| Q \right|=\sqrt{\left(\Delta-\frac{\alpha^2}{\Delta-\delta}\right)^2-\frac{1}{4}} ,
\end{equation}
where $\alpha=\Omega_{\scr{P}}/(2\Omega_{\scr{B}})$ characterizes the relative 
strength of the Bragg and polaritonic couplings. 
Expression~(\ref{spectrum}) is 
our main result. It clearly demonstrates that the Bragg 
and polaritonic dispersion relations {\em compete}
with each other. 

Consider for simplicity the case of exact resonance, 
{\em i.e.} $\delta=0$. 
It is seen from Eq.~(\ref{spectrum}) that in the absence of 
light-matter coupling ($\alpha=0$), the first term in the 
brackets gives rise to the
conventional PBG. It is also seen that with
increasing $\alpha$ (or, $\Omega_{\scr{P}}$) the decay
length, $\mbox{Im}\;Q^{-1}$, increases, and for sufficiently small  
$\Delta$ we find 
%%%%%%%%%%%%%%%%%%%%%%%%%%%%%%%%%%%%%%%%%%%%%%%%%%%%%%%
%%%%%%%%%%%%%%%%%%%%%%%%%%%%%%%%%%%%%%%%%%%%%%%%%%%%%%%
\parbox[t]{8.65cm}{
\psfig{figure=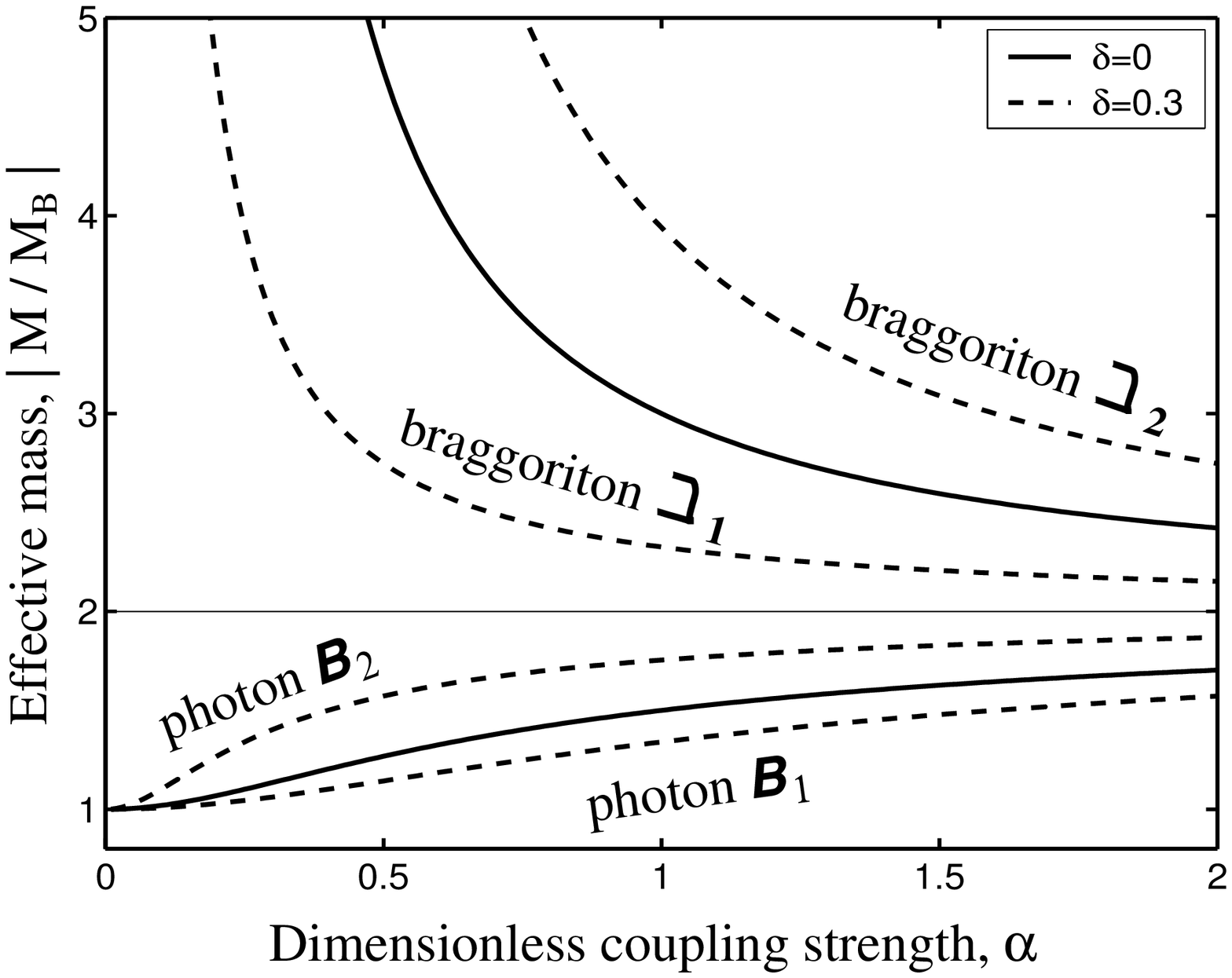,width=8.65cm}

\vskip2mm

{
\small{\bf Fig. 2}
The effective mass for various excitations
(in units
of ``free'' Bragg mass 
$M_{\scr{$\mb{\cal B}$}}=
M_{\scr{$\mb{\cal B}_1$}}=
M_{\scr{$\mb{\cal B}_2$}}$ at $\alpha=0$) is plotted
vs. coupling strength: solid lines are for
$\delta=0$ where $M_{\scr{$\mb{\cal B}1$}}=
M_{\scr{$\mb{\cal B}2$}}$,
$M_{\scr{$\mbox{\eightta a}_1$}}=
M_{\scr{$\mbox{\eightta a}_2$}}$; dashed lines are for
$\delta=0.3$.
}

\vskip4mm

}
%%%%%%%%%%%%%%%%%%%%%%%%%%%%%%%%%%%%%%%%%%%%%%%%%%%%%%%%
%%%%%%%%%%%%%%%%%%%%%%%%%%%%%%%%%%%%%%%%%%%%%%%%%%%%%%%%
that
$Q$ becomes {\em real}. This manifests the emergence of the
novel {\em allowed} photonic states, or braggoritons, inside
the PBG (see Fig.~1). The braggoriton branches in the excitation dispersion 
relations are
described by the operators $\hat{\mbox{\ta a}}_1$ and 
$\hat{\mbox{\ta a}}_2$. 
They occupy the frequency ranges
%\Bigl(0,\pm\frac{1}{4}[+\frac{1}{2}\sqrt{\frac{1}{4}+2\alpha^2})\}$, 
$\Delta=
\Bigl[0, \pm\frac{1}{4}\Bigl(\sqrt{1+16\alpha^2}-1
\Bigr)\Bigr]$. 
For small $\alpha$ ($\alpha\ll 1$), the braggoriton frequency
interval reduces to $\Bigl(0, \pm2\alpha^2\Bigr)$.
We note that due to the finite $\alpha$ value
the Bragg gap broadens. Namely,
the band edges of the branches described by the
operators $\mb{\hat{\cal B}}_1$, 
$\mb{\hat{\cal B}}_2$ are respectively given for $\delta=0$ by
$\Delta=\pm\frac{1}{4}\Bigl(\sqrt{1+16\alpha^2}+1\Bigr)$ 
(compare to $\Delta=\pm1/2$ for $\alpha=0$).
The dispersion relations $\Delta(Q)$ calculated using 
Eq.~(\ref{spectrum}) are shown in Fig.~1(a) for 
different values of $\alpha$ in the case of exact resonance
$\omega_{\scr{B}}=\omega_{\scr{T}}$, or $\delta=0$.

Moderate frequency detuning $\delta\neq 0$
does not qualitatively change the above picture as
seen in Fig.~1(b). The major effect of frequency detuning is that 
 the braggoritonic branches $\mbox{\ta a}_1$, $\mbox{\ta a}_2$ 
acquire an asymmetry since 
they are  ``pinned'' by $\omega_{\scr{T}}$. 
Fig.\,1 also shows that the Bragg-like
photonic branches 
$\mb{\cal B}_1$, $\mb{\cal B}_2$ are affected by coupling or 
detuning only weakly.

To quantitatively describe the braggoriton dispersion relation,
we consider two  characteristics: 
({\em i}) dimensionless effective mass, $M$ near the band edges that is
defined from Eq.~(\ref{spectrum}) by the relation,
\begin{equation}
\Delta\simeq \Delta_{Q=0}+\frac{Q^2}{2M},
\end{equation}
and ({\em ii}) the density of states, $N(\Delta)$.
Expanding Eq.~(\ref{spectrum}) in $Q$
yields the following effective
masses for the braggoritons \\
%%%%%%%%%%%%%%%%%%%%%%%%%%%%%%%%%%%%%%%%%%%%%%%%%%%%%%%
%%%%%%%%%%%%%%%%%%%%%%%%%%%%%%%%%%%%%%%%%%%%%%%%%%%%%%%
\parbox[t]{8.65cm}{
\psfig{figure=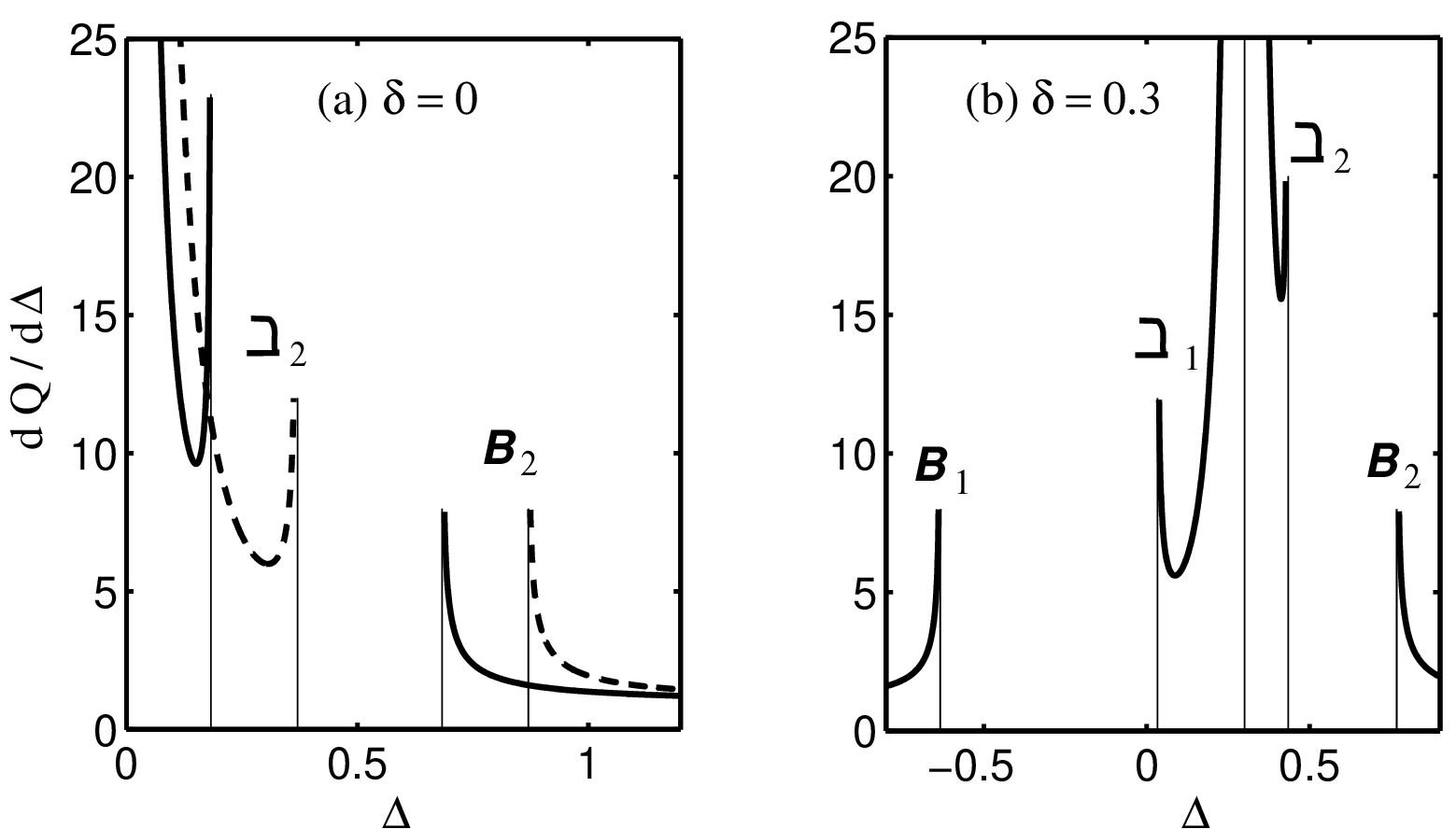,width=8.65cm}

\vskip2mm

{
\small{\bf Fig. 3}
Density of states for mixed photonic-medium excitations. 
Left panel is for the
symmetric case ($\delta=0$), for
$\alpha=0.35$ (solid line) and $\alpha=0.57$ (dashed line). Right panel is
for $\delta=0.3$ and $\alpha=0.35$. The thin solid vertical lines indicate the
band edges.
}

\vskip4mm

}
%%%%%%%%%%%%%%%%%%%%%%%%%%%%%%%%%%%%%%%%%%%%%%%%%%%%%%%%
%%%%%%%%%%%%%%%%%%%%%%%%%%%%%%%%%%%%%%%%%%%%%%%%%%%%%%%%
%
\begin{eqnarray} \label{pimass1}
M_{\scr{$\mbox{\eightta a}_1$}}&=&\left[1-\frac{1-2\delta}{\sqrt{\left
(1-2\delta
\right)^2+16\alpha^2}}\right]^{-1}, \\ \label{pimass2}
M_{\scr{$\mbox{\eightta a}_2$}}&=&
-\left[1-\frac{1+2\delta}{\sqrt{\left(1+2\delta
\right)^2+16\alpha^2}}\right]^{-1}.
\end{eqnarray}
These masses are plotted in Fig.~2 versus the coupling strength
 $\alpha$.
For $\alpha\rightarrow 0$, we have $M_{\scr{$\mbox{\eightta a}_1$}}$,
$M_{\scr{$\mbox{\eightta a}_2$}}\rightarrow\infty$, 
reflecting the fact that 
at $\alpha=0$ the braggoritons reduce to dispersionless
medium  excitations that are not coupled to light.
With the light-matter interaction switched on, the braggoriton
effective mass rapidly decreases 
(the width of the in-gap branches increases).

The one-dimensional density of states $N(\Delta)$ is given from
Eq.~(\ref{spectrum}) by:
\begin{eqnarray}\label{density}\nonumber
& &N(\Delta)\propto \frac{dQ}{d\Delta}= \\
& &\frac{\Delta-
\ds\frac{\alpha^2}{\Delta-\delta}}{\sqrt{
\left(\Delta-
\ds\frac{\alpha^2}{\Delta-\delta}\right)^2-\ds\frac{1}{4}}}\;\left(1+
\frac{\alpha^2}{(\Delta-\delta)^2}\right),
\end{eqnarray}
and shown in Fig.~3 for different values
of $\alpha$ and detuning, $\delta$. The density of
states of the braggoritonic branches inside the gap exhibits conventional
1D square-root singularities at the band edges 
$\Delta=\delta$ and 
$\Delta=\frac{1}{2}\Bigl[\delta\mp\frac{1}{2}\pm\sqrt{(\delta\pm
\frac{1}{2})^2+4\alpha^2}\Bigr]$.

As mentioned above, the upper and lower
Bragg-like photonic 
branches,  $\mb{\cal B}_1$, $\mb{\cal B}_2$,
are only 
slightly affected by the coupling and/or detuning. In particular their
effective masses,
\begin{eqnarray}
M_{\scr{$\mb{\cal B}_1$}}&=&-
\left[1+\frac{1+2\delta}{\sqrt{\left(1+2\delta
\right)^2+16\alpha^2}}\right]^{-1}, \\
M_{\scr{$\mb{\cal B}_2$}}&=&
\left[1+\frac{1-2\delta}{\sqrt{\left(1-2\delta
\right)^2+16\alpha^2}}\right]^{-1}, 
\end{eqnarray}
change only by a factor of 2 as $\alpha$ varies from zero
to infinity (see Fig.\,2).

\section{Intragap localized states}

We now turn our attention to the 
localized photonic states caused by a phase-slip like defect. Note that
in the absence of the polarizable medium, a structure with 
one-dimensional modulation~(\ref{epsilon}) 
of the dielectric function can be viewed as a distributed
feedback resonator\cite{PBG} first considered by Kogelnik and 
Shank\cite{Kogelnik} in
1972. Later it was realized that a phase slip\cite{slip} in the
modulation
\begin{equation}
\varepsilon(z)=\varepsilon_0+\delta\varepsilon\cos(\sigma z+\phi(z)),
\end{equation}
where
\begin{equation} \label{phaseslip}
\phi(z)=\left\{\begin{array}{ll}
\phi_1,\hskip3mm & z<0 \\
\phi_2, & z>0
\end{array}\right.,
\end{equation}
results in a localized state inside the PBG. Within the second
quantization formalism of Sec.~II, the emergence of such a state
can be established as follows. Consider the eigenstate
annihilation operators
of the Hamiltonian~(\ref{Hph})
\begin{equation} \label{operators}
\left(\begin{array}{c}
\hat{\beta}_1 \\
\hat{\beta}_2
\end{array}\right)=
\left(\begin{array}{cc}
\cos\theta & -\sin\theta\;e^{i\phi} \\
\sin\theta\;e^{-i\phi} & \cos\theta
\end{array}\right)\; \left(\begin{array}{c}
\hat{a}_{\rightarrow} \\
\hat{a}_{\leftarrow}
\end{array}\right).
\end{equation}
It follows from~(\ref{operators}) that the absolute value, $\lambda$, 
of the amplitude ratio of
the left and right propagating waves constituting the eigenstates 
$\beta_1$, $\beta_2$ is either $\lambda=\tan\theta$, or 
$\lambda=\tan^{-1}\theta$.
These expressions are actually equivalent with appropriate choice
of the sign of square root:
\begin{equation}\label{lambdatan}
\lambda_{\pm}(\Delta)=2\left(\Delta\pm\sqrt{\Delta^2-\ds\frac{1}{4}}\;\right),
\end{equation}
where we used the definition~(\ref{costheta}) of the rotation angle $\theta$.

In the presence of a phase slip~(\ref{phaseslip}) the continuity condition
at $z=0$ reads
\begin{equation} \label{lambdacond}
\lambda_{-}(\Delta) e^{-i\phi_2}=\lambda_{-}^*(\Delta) e^{-i\phi_1}.
\end{equation}
As is well known\cite{slipsolution} 
Eq.~(\ref{lambdacond}) has a unique in-gap
solution, $\Delta'$, for an arbitrary phase discontinuity $\phi_1-\phi_2$,
\begin{equation} \label{ingap}
\Delta'=\cos\chi,
\end{equation}
where
\begin{equation} \label{chidef}
\chi=\left\{\begin{array}{ll}
\ds\frac{\phi_1-\phi_2}{2}+\pi,\hskip3mm & -\pi<\phi_1-\phi_2<0, \\
\ds\frac{\phi_1-\phi_2}{2}, & 0 < \phi_1-\phi_2 <\pi.
\end{array}\right.
\end{equation}
%
%%%%%%%%%%%%%%%%%%%%%%%%%%%%%%%%%%%%%%%%%%%%%%%%%%%%%%%
%%%%%%%%%%%%%%%%%%%%%%%%%%%%%%%%%%%%%%%%%%%%%%%%%%%%%%%
\parbox[t]{8.65cm}{
\psfig{figure=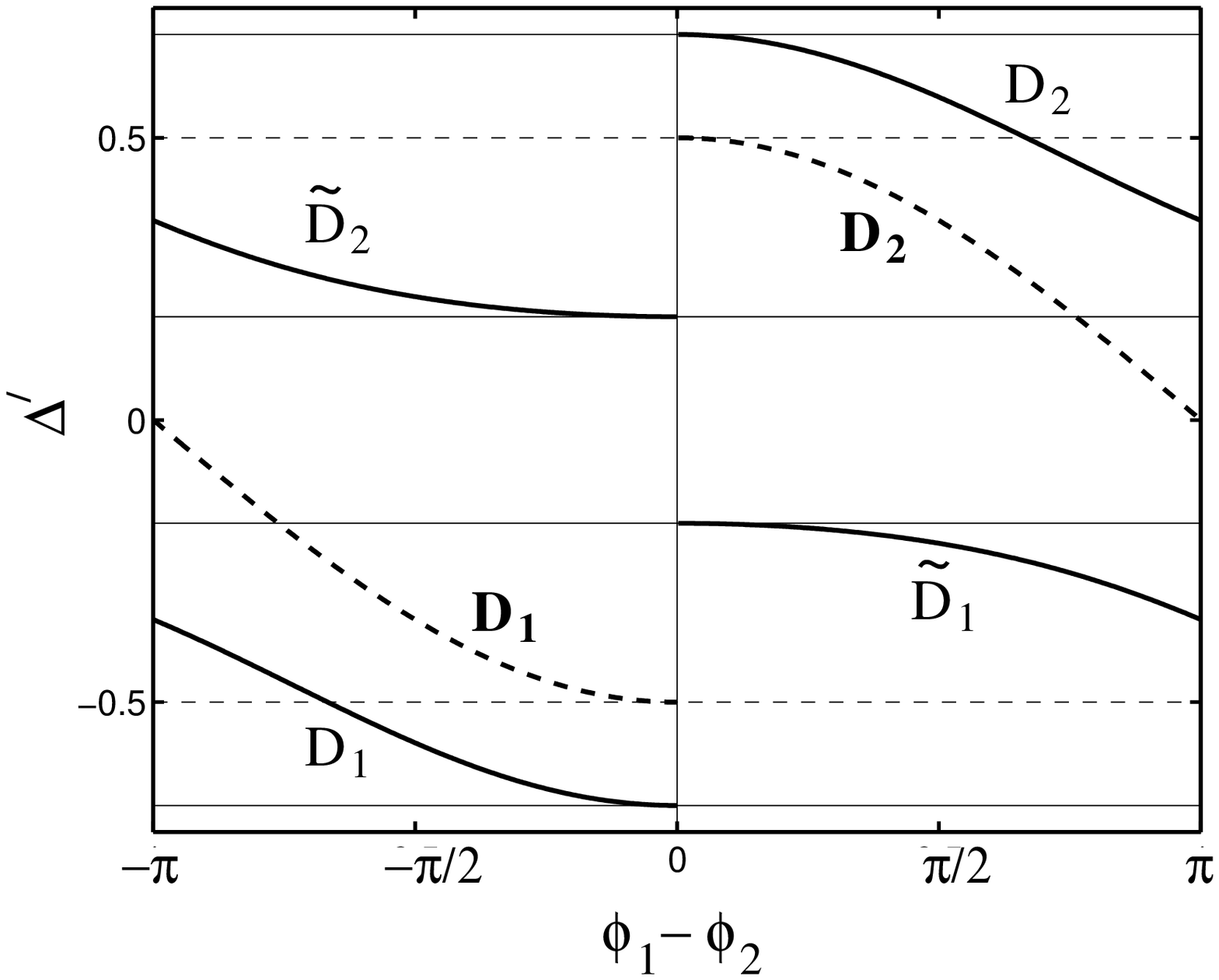,width=8.65cm}

\vskip2mm

%{
\small{\bf Fig. 4}
The frequencies of localized intragap states vs. the phase slip
magnitude: 
{\bf D}$_1$, {\bf D}$_2$ 
are defect levels inside the Bragg gap
($\alpha=0$, dashed line); $D_1$, $\tilde{D}_1$ are defect levels
inside the lower subgap, and $D_2$, $\tilde{D}_2$ are defect levels
inside the upper subgap ($\delta=0$, $\alpha=0.35$, solid line). 
Thin solid and dashed lines represent the
band edges of the two forbidden gaps 
and of conventional Bragg gap (in the absence
of coupling), respectively.
%}

\vskip4mm

}
%%%%%%%%%%%%%%%%%%%%%%%%%%%%%%%%%%%%%%%%%%%%%%%%%%%%%%%%
%%%%%%%%%%%%%%%%%%%%%%%%%%%%%%%%%%%%%%%%%%%%%%%%%%%%%%%%
Generalization of the above consideration to include the
polarizable medium is straightforward. It reduces to 
the following modification of the parameter $\lambda$ in 
Eq.~(\ref{lambdacond}):
\begin{eqnarray}\label{lambda}\nonumber
& &\lambda_{\pm}(\Delta,\alpha,\delta)= \\
& &2\left(\Delta-
\frac{\alpha^2}{\Delta-\delta}\pm
\sqrt{\left( \Delta-
\frac{\alpha^2}{\Delta-\delta}\right)^2-\frac{1}{4}}\;\right),
\end{eqnarray}
Then  condition~(\ref{lambdacond}) yields the gap state solution, $\Delta'$
\begin{equation} \label{slipdelta}
\Delta'=
\frac{1}{2}\left[\delta+\frac{1}{2}\cos\chi\pm\sqrt{\left(\delta-\frac{1}{2}
\cos\chi\right)^2+4\alpha^2}\;\right],
\end{equation}
where $\chi$ is defined by Eq.~(\ref{chidef}).
For $\alpha=0$ we return to the in-gap state~(\ref{ingap}). Remarkably, 
we note 
that for nonzero coupling parameter $\alpha$, when the Bragg gap
is divided into two sub-gaps as in Fig.~1, the two 
values of $\Delta'$ determined 
by Eq.~(\ref{slipdelta}) are located in {\em each} of the corresponding
sub-gaps.
In Fig.~4 the frequencies, $\Delta'$ of the localized in-gap states
are shown versus the magnitude $\phi_1-\phi_2$ of the phase-slip
for zero detuning $\delta=0$.

\section{Discussion}

%We studied properties of one-dimensional photonic crystal filled with 
%polarizable medium. We have shown that in the presence of both Bragg 
%scattering and light-matter coupling, 
%new branches of propagating excitations appear inside the Bragg
%gap. We have analyzed the dispersion law of these excitations.

Here we discuss the connection of the present work to 
the earlier related studies.

%We also studied the effect of the phase slip in the modulation of the 
%dielectric function and demonstrated that for arbtitrary magnitude of 
%the phase jump, a localized state exists in each of the two forbidden
%sub-gaps formed in the system.

The band structure of a photonic crystal with frequency dependent
dielectric function was recently studied numerically in 
Ref.\onlinecite{Mpolariton}, using the plane-wave method. 
In this work the photonic crystal was modeled as a 
two-dimensional array of GaAs rods. Frequency dispersion was
introduced through the transverse-longitudinal
splitting of the optical phonons. The authors\cite{Mpolariton} 
(see also Ref.\onlinecite{china}) observed that
numerous branches of the band structure calculated for
$\omega_{\scr{B}}\equiv \omega_{\scr{T}}$ become almost
dispersionless at frequencies close to the frequency $\omega_{\scr{T}}$.

In the context of the present work, this weakening of dispersion
can be understood from Eqs.~(\ref{pimass1}), (\ref{pimass2}) that describes 
the effective masses
of the braggoritonic branches $\mbox{\ta a}_1$, $\mbox{\ta a}_2$. 
These masses  
rapidly increase as the coupling parameter $\alpha$ decreases.

Signatures of braggoritonic excitations studied in detail
in the present paper can
be also found in the numerical calculations of
Ref.\onlinecite{Spolariton}. In that work, transmission spectra of a
photonic crystal identical to that of Ref.\onlinecite{Mpolariton} were
calculated within the transfer-matrix formalism.  {\em Two} minima in
the transmission spectra were found instead of the usual {\em single} minimum
that is caused by the Bragg diffraction 
in non-dispersive photonic crystal. In
light of the theory developed in the present work, these two minima can be
identified with the {\em two forbidden subgaps} in the excitation
spectrum (Figs.~1 and~3). As we have demonstrated (Fig.~3), 
in the presence of 
light-matter coupling there are two spectral regions with zero
density of states. Correspondingly, the transmission coefficient
within these frequency regions must be low if the sample is 
sufficiently thick.

A very different realization of periodic polarizable structures
was the subject of  extensive theoretical studies during the last 
decade.\cite{Ivchenko,TETM}
The structures are multiple quantum wells separated 
by wide-gap semiconductor barriers. The width, $d$, 
of each barrier was assumed to
be close to $\lambda_0/2$, where $\lambda_0$ is the wavelength
corresponding to the intra-well exciton resonance 
frequency, $\omega_{\scr{T}}$. The condition $d\approx\lambda_0/2$
implies that the Bragg frequency, $\omega_{\scr{B}}$ is close to 
$\omega_{\scr{T}}$.
Since the quantum wells with strong frequency
dispersion at $\omega\sim \omega_{\scr{T}}$ had thickness much smaller
than $d$,  then a real Bragg gap in the 
structures\cite{Ivchenko,TETM} was lacking. However, under the
condition $\omega_{\scr{T}}\equiv \omega_{\scr{B}}$ the dispersion 
law of light propagating along the principal axis was     
shown to have a gap within a frequency range 
$\left|\omega-\omega_{\scr{T}}\right|
=\left(2\Gamma_0\omega_{\scr{T}}/\pi\right)^{1/2}=\Omega^{\scr{(eff)}}$.
Here $\Gamma_0$ denotes
the radiative rate for an exciton in a single well.
In other words, $\Omega^{\scr{(eff)}}$
plays the role of the ``effective'' Bragg gap in the 
structures.\cite{Ivchenko,TETM}
Remarkably, a  physical picture completely analogous 
to the multiple-quantum-well structures emerged from consideration 
of an optical lattice formed by laser-cooled atoms.\cite{Phillips} 
Correspondingly, the light dispersion relation  derived in 
Ref. \onlinecite{Phillips} has the same form as in~ 
Refs.~\onlinecite{Ivchenko},~\onlinecite{TETM}.
In conclusion we note that as was recently pointed out\cite{Deych}, a detuning 
$(\omega_{\scr{B}}-\omega_{\scr{T}})\sim \Omega^{\scr{(eff)}}$ gives rise to a
band of propagating states within the ``effective'' Bragg gap
of the multiple-quantum-well structures.

{\bf Acknowledgments:}
This work was supported by NSF grant DMR 9732820, the
Petroleum Research Fund under grant ACS-PRF \#34302-AC6, and the Army
Research Office.
\unsethebrew

\end{multicols}

\end{document}